\documentstyle [prl,aps]{revtex}
\input epsf
\begin{document}

\draft
\title{Constraints on Exotic Heavily Ionizing Particles 
from the Geological Abundance of Fullerenes}

\author{
J.I. Collar$^{a,b,*}$ and K. Zioutas$^{c}$
}
\address{ 
$^{a}$Groupe de Physique des Solides (UMR CNRS 75-88), Universit\'e Paris 7, 
2 Pl. Jussieu, Paris 75251, France\\
$^{b}$CERN, EP Division, CH-1211 Geneve 23, Switzerland\\
$^{c}$Physics Department, University of Thessaloniki, GR-54006 
Thessaloniki, Greece
}
\wideabs{
\maketitle
\begin{abstract}
\widetext
The C$_{60}$ molecule exhibits a remarkable stability that leads 
to its survival in ancient carbonaceous rocks 
initially subjected to the elevated temperature requisite for its 
formation. Elementary particles having a large electronic stopping power 
can similarly form C$_{60}$ and higher fullerenes in their wake. Combined, 
these two features point at the possibility of using the C$_{60}$ presence 
in selected bulk geological samples as a new type of 
nuclear track detector, with applications in astro-particle physics. 
\end{abstract}

\pacs{PACS number(s): 
95.55.Vj, 95.35.+d, 14.80.-j, 78.30.Na, 61.82.Pv\\
$^{*}$~Corresponding author. E-mail: Juan.Collar@cern.ch}
}

\narrowtext
Sixty carbon atoms, arranged spherically as if on the surface of a soccer ball, 
form the C$_{60}$ molecule (Buckminsterfullerene). Of all possible formations, 
this peculiar one has the highest symmetry, lowest energy level, no dangling 
bonds and is therefore chemically inert much like a noble gas. This leads to 
an extraordinary photochemical and thermal stability that fueled the 
expectations for its abundance in the galaxy \cite{Kroto} soon after its 
discovery. First searches for C$_{60}$ in meteorites and the interstellar 
medium turned up negative \cite{Heymann1,DeVries}; the presence of hydrogen as an 
abundant co-reactant in stellar atmospheres was recognized
as an inhibiting factor in its production 
\cite{DeVries}. In spite of this, there is now a growing evidence for C$_{60}$ 
in the skies \cite{Foing}. On Earth, the confirmation of a natural 
C$_{60}$ presence followed a similar chronicle: Only particular geological 
samples having once withstood the high temperature conditions 
requisite for C$_{60}$ formation have so far shown this evidence. This is the 
case of pre-cambrian Shungite \cite{Buseck}, the most highly 
methamorphosed coal known, or 
Fulgurite \cite{Daly}, the glassy product of a lighting strike. Of particular 
interest is its presence in shock-produced rubble in the 1.85 billion-year-old 
Sudbury impact crater \cite{Becker1,Becker2} and in samples from 
Cretaceous-Tertiary (K/T) 
boundary clay seams and coals \cite{Heymann2}. In the first case, atomic 
helium encapsulated in the fullerene exhibits an extraterrestrial 
$^3$He/$^4$He ratio, pointing at an origin previous 
to the impact itself \cite{Becker2}. The high concentration of fullerenes 
(1-10 ppm) in 
these samples illustrates the long-term survival of the molecule, 
and the ability to 
retain extraterrestrial He its endurance of extreme conditions such as those 
accompanying crater formation. The C$_{60}$ presence in 
concentrations typical of soots from common flames in K/T samples, but not 
above or below the boundary, is hypothesized to arise 
from the raging of post-impact
sooting wildfires \cite{Heymann2}.

In the laboratory, fullerenes have been produced by laser ablation, in carbon 
arcs, by combustion of benzene and lately by ion beam irradiation 
\cite{Chadderton1,Fink}. C$_{60}$ and other polycyclic compounds 
remain in latent tracks of energetic ions in carbonaceous matter, 
albeit a minimum of few eV/\AA$^3$
in transferred electronic energy is needed for this process
\cite{Chadderton1,Gamaly,Bitensky}, so as to permit the completion of
chemical formation 
before the onset of cooling in the particle's hot plasma aftertrack. 
Given the apparent scarcity of fullerenes in common geological 
samples, yet 
its demonstrated ability to survive over long periods, it is natural to wonder if 
hypothetical highly-ionizing and deeply-penetrating elementary particles 
such as magnetic monopoles, 
\mbox{nuclearites \cite{Alvaro},} $Q$-balls \cite{Coleman}, etc.,
shouldn't have left a detectable cumulative signature in 
ancient carbonaceous rocks. In this Letter, recent models and 
measurements of C$_{60}$ production under particle irradiation are 
employed to estimate its 
yield from such cosmic exotica, comparing it with that from 
common terrestrial sources of radiation. This geo-chemical method is 
appraised judging from the sensitivity of present-day chromatographic 
techniques and current efficiency in C$_{60}$ extraction from bulk minerals.

A condition for the applicability of this technique is that of good 
fullerene preservation over the age of the rock. There are 
several precautions to be taken in this respect for sample selection, a 
situation similar to that of mica searches \cite{Daniel,Beauford} for Weakly Interacting 
Massive Particles (WIMPs) and 
monopoles, where the thermal history of the mineral must be 
such that annealing of latent etch-tracks is minimized. While C$_{60}$ 
integrity under impact and static pressure in the 
upper layers of the Earth's crust seems assured (it is stable up to 130-170
kbar \cite{Yoo}), there are factors such as exposure to O$_3$ 
\cite{Chibante2} and UV light \cite{Taylor} to be 
controlled. An environment low in ozone is imperative since the destruction of 
fullerenes is five orders of magnitude faster in the presence of O$_3$ than 
O$_2$. In fact, the Shungite of Ref. \cite{Buseck} is thought to have been 
protected from O$_3$ oxidation by the coal in which it was contained 
\cite{Chibante2}. The concentration of hydrogen and sulfur in the rock 
should also be a criterion for selection, the second preventing C$_{60}$ oxidation 
when present 
as sulfide-silicates \cite{Becker1}. The thermal decomposition of 
fullerenes in solid phase, another determining factor, has been measured \cite{Leifer}: 
The remaining fullerene fraction after a time $t$ at 
temperature $T$ is 
$\exp\left(-k\left(T\right)t\right)$, 
where
$k\left(T\right)=1.24\times10^{9}\exp\left(\frac{-\tau}{T}\right)  
\mathrm{s}^{-1}$ and
$\tau=3.2\times10^{4}~ \mathrm{K}$, i.e.,
a 90\% survival probability over $10^{9}$ years even if subjected to a 
constant 250$^{\circ}$C. Finally, fullerenes exhibit a very good stability 
towards intense irradiation with high-energy electron pulses and gamma 
rays \cite{Maity}, guaranteeing their resistance to scarce
environmental minimum ionizing radiations. In view of these remarkable 
signs of immunity, the ppm presence or not of fullerenes in 
ancient rocks seems to be less a question of long-term stability and more 
of sample origin, as experimental findings confirm. If exposure to certain 
elementary particles over the lifetime of the sample is an 
abundant source of fullerenes, these should endure to evidence the particle's 
crossing. 

It is the ability to efficiently leach fullerenes 
from bulk mineral and to detect them that ultimately limits this 
method. In this 
Letter the state-of-the-art in High-Pressure Liquid 
Chromatography (HPLC) is conservatively adopted as the detection 
reference.
Similarly, we do not assume a leaching efficiency far beyond what is  
accomplished nowadays, i.e., that O(1) kg samples can be nearly completely 
demineralized \cite{Heymann2,Heymann4} and reduced to $\sim$0.1 ml of 
liquid containing all organic substances of interest. In essence, this procedure 
consists of pulverization followed by repeated treatment with HCl/HF to 
create a C-rich residue, refluxing of this in a small volume of toluene (a good 
solvent for aromatic compounds), centrifugation and/or 
filtration, and concentration of the volume to 0.1-1 ml by evaporation 
under reduced pressure \cite{Heymann2,Heymann4,Wolbach}.
The recovery efficiency for thus-treated fullerene-spiked mineral 
samples has been shown to be $\sim$90\% \cite{Heymann4}. At a reasonable 
signal-to-noise ratio, HPLC is currently able to spot $\sim\!\!10^{-10}$ g of 
fullerene in 0.1 ml toluene aliquots \cite{Fink,Heymann2}: this is the realistic 
figure adopted below to estimate the present particle detection ability.

The production of fullerenes in the wake of an energetic particle follows a 
sequence of events \cite{Chadderton1,Gamaly,Bitensky,Fink} that begins 
with the destruction of all molecular bonds in the track core due to the 
formation of a short-lived zone of highly excited neutral target atoms by 
the electronic energy transfer from the projectile. This is followed by the 
nucleation of one-dimensional C-C complexes via random ``sticky'' 
collisions of this gas, coalescing into two-dimensional pentagonal and 
hexagonal rings that in turn form closed 3-d complexes, and among those 
fullerenes. The concept of a threshold in electronic stopping 
power for fullerene formation is interchangeable with that of a minimum 
track temperature above which condensation is facilitated \cite{Bitensky}. 
The dependence of the track temperature 
$T\left(r,t\right)$ on the axial distance $r$ from 
the particle path and time $t$ is \cite{Mozumder}:
\begin{equation}
T\left( r,t\right) = \frac{T_0}{1 + \left( 4D\,t / r^{2}_{0} \right) }
\exp \left\{ - \frac{\left( r / r_{0} \right) ^{2}}{1 + \left( 4D\,t / r^{2}_{0} 
\right)} \right\},
\end{equation}
where $D$ is the thermal diffusivity of the medium and $r_{0}\!\!\sim$15 
\AA ~is the core radius, which delimits a cylindrical region of
plasma-like conditions and highest energy density \cite{Mozumder}. 
The initial track temperature is given by
$T_{0} = \alpha S_{e}/(\pi \rho C_{V} r^{2}_{0})$,
with $\rho$ as the density, C$_{V}$ the specific heat, S$_{e}$ the electronic stopping 
power and $\alpha\!\!\sim$0.1 as the fraction of energy going into thermal 
excitation of the Maxwellian gas of carbon atoms. 
Leaving aside a dependence on the probability of sticky C-C collisions and 
on the average density of the gas (both temperature-independent to first 
approximation \cite{Bitensky,Creasy}), the C$_{60}$ yield is  proportional to:
\begin{equation}
Y_{60} \propto \int\!\!\!\int\!\!\!\int \Omega \left( r,T \right) ~2 \pi r~dr~dz~dt
\end{equation}
($z$ denotes the direction along the trajectory). This expression gives the time-integrated
 volume of material available for fullerene condensation under the 
favorable conditions described by $\Omega \left( r,T \right)$, namely:
\begin{equation}
\Omega \left( r,T \right) = \left\{
\begin{array}{ll}
1 \mathrm{\ for\ } \emph{T}_{min} < \emph{T} \left( \emph{r,t} \right) < \emph{T}_{max}, & 0 
\mathrm{\ elsewhere}\\
0 \mathrm{\ for\ } \emph{r} < \emph{r}_{60}, & 1 \mathrm{\ elsewhere,}
\end{array}
\right.
\end{equation}
where the minimum temperature for fullerene formation is estimated at 
$T_{min}\!\!\sim$750-800 K in studies of laser ablation of polymers 
\cite{Creasy,Gamaly}, and  unconfined C$_{60}$ is seen to readily disintegrate 
at $T_{max}\!\!\sim$6000 K. The second condition in 
$\Omega \left( r,T \right)$ imposes that the heated volume must be able to 
physically accommodate the C$_{60}$ cage within; $r_{60}$ should then naively 
correspond to the molecule's radius, 3.6 \AA. A more adequate 
phenomenological value $r_{60}\!\!\sim$9 \AA \cite{Bitensky} is adopted here.

In the present analysis, $T \left( r,t \right)$ must be evaluated numerically to compute
$Y_{60}$ for different types of radiation. These yields 
are in arbitrary units until a normalization point is applied; of the handful of 
recent irradiations of carbonaceous materials with heavy ions, the results of Fink 
\emph{et al.} \cite{Fink} using 3 GeV U$^{20+}$ on thin plates of 
sucrose ($S_{e}\!\!\sim\!\!2.0\times 10^{4}$keV/$\mu$m, ion range $\sim$160$\mu$m) 
are particularly helpful for 
this purpose, since the total dissolution of sugar in HPLC reagents guarantees an 
excellent C$_{60}$ recovery. The found value of 84$\pm$31 C$_{60}$ molecules 
per incident ion should then reflect the actual $Y_{60}$ closely. 
Eq.(2) can be used to assess $Y_{60}$ had the target been instead a 
carbonaceous rock: The different stopping powers \cite{TRIM} and parameter values
for sucrose ($C_{V}\!\!\sim\!\!1.2\times10^{3}$
J/(kg K), $D\!\!\sim\!\!2.9\times10^{-7}$ 
m$^{2}$/s, $\rho\!\!\sim\!\!1.6 \times 10^3$ kg/m$^{3}$) and 
coal-like samples ($C_{V}\!\!\sim\!\!700$ J/(kg K), $D\!\!\sim\!\!
1.8 \times 10^{-7}$ m$^{2}$/s, $\rho\!\!\sim\!\!2 \times 10^3$ kg/m$^{3}$) 
translate into an enhancement in $Y_{60}$ by a factor 3.8. 
The nearly constant $S_{e}$ over the initial 90\% of the 
ion range in both situations allows to conclude that, 
to a good approximation, $\frac{\partial Y_{60}}{\partial z}\!\simeq\! 
2.5\times 10^{-4}$ C$_{60}$/\AA ~at 
$S_{e}\!\!\simeq\!\!2.4\times 10^{4}$ keV/$\mu$m for C-rich rocks 
(\emph{the 
units of $S_{e}$ and value of $\rho$ are maintained 
tacitly hereafter}). This 
normalization point anchors our calculation of $\frac{\partial 
Y_{60}}{\partial z}$ vs $S_{e}$, shown in Fig.1.
{\frenchspacing Fink \emph{et al.} observed} that C$_{60}$ 
production was a highly inefficient process (36 MeV invested per C$_{60}$ when 
the molecule's formation energy is 444 eV). An 
interpretation \cite{Fink,Finkmo} is that the carbon atoms are highly 
volumetrically diluted in sucrose (one in four), reducing the 
probability of C-C collisions; this is aggravated by the presence of 
hydrogen, which 
can lead to aromatic hydrocarbons rather than fullerenes as stable end 
products \cite{DeVries}. There should then be an additional enhancement 
in $Y_{60}$ for rocks primarily composed by C. For these the 
overall normalization 
of Fig.1 is probably largely conservative and not a substitute for 
calibrations using ions of stopping power similar to that of the particle of interest 
(calibrations that should also determine the leaching efficiency for 
an specific rock type).
\begin{figure}[tbp]
\epsfxsize = \hsize \epsfbox{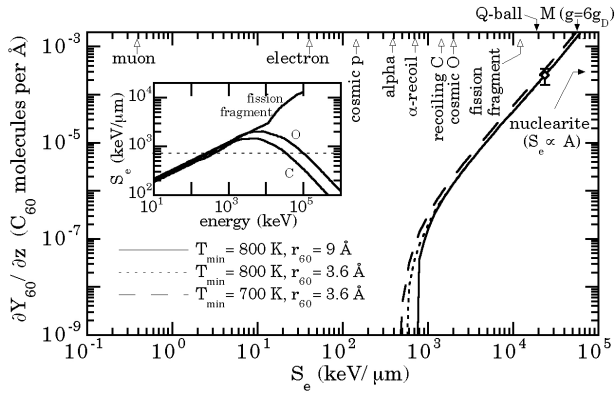}
\caption{Estimated C$_{60}$ yield per unit path length vs particle electronic
stopping power, in coal-like rocks. The single point 
corresponds to the overall normalization based on [13]. The dependence 
on $T_{min}$ and $r_{60}$ is shown; similar variations of $r_{0}$ and
$T_{max}$ have a comparatively negligible effect. The slope at high 
$S_{e}$ is in good agreement with [12]. The \emph{maximum}
$S_{e}$ for common natural sources of radiation is indicated (white 
arrows) and 
illustrated in the insert for those above the fullerene 
production threshold $S_{e}\sim$750 keV/$\mu$m.} 
\end{figure}
Do natural backgrounds give rise to a significant fullerene presence 
in shallow-mined carbonaceous rocks? Isolated minimum 
ionizing radiations cannot play a role (Fig. 1). However, recoiling 
carbon nuclei with energies $\geq30$ MeV ($Y_{60}\!\!\simeq$0.1) can be induced by 
underground muons, albeit the cross section for this 
process \cite{Jackson} is small at $<\!\!10^{-3}$mb. With a muon flux 
$10^{-3}\mathrm{cm}^{-2}\mathrm{s}^{-1}$ at a 
modest 150m depth, these recoils should produce a meager 
$<\!10^{-21}\mathrm{g~C}_{60}/\mathrm{kg~rock/y}$. A comparable 
contribution, if any, is expected from O(1)MeV recoils 
($Y_{60}\!\!\simeq\!\!10^{-3}$) 
originating in elastic scattering of the $2\times10^{-7}\mathrm{cm}^{-2}\mathrm{s}^{-1}$
fast neutron flux typical below this depth. Similarly, recoiling daughters from 
(n,p),(n,$\alpha$),($\alpha$,p),($\alpha$,n) reactions in C can be 
shown to contribute negligibly, and the same applies to the scarce 
penetrating medium- and large-mass cosmic ray nuclei. The biggest offenders 
are, by far, 
spontaneous fission fragments ($Y_{60}\!\!\simeq\!\!5$) which result in 
$2.8\times10^{-18}\mathrm{g~C}_{60}/\mathrm{kg~rock/y/ppm}^{238}U$, 
making a low U concentration necessary if ancient samples of age 
$\sim$1 Gy are to be utilized (US coals exist with $<$0.02 ppm 
U). This is again reminiscent of WIMP mica searches, where 
$\leq10^{-4}$ppm U are desirable.

The central result of our analysis is that heavily ionizing particles energetic enough 
to maintain a constant $S_{e}>10^{3}$ 
through the Earth's upper crust should produce at least
$\sim2.05\times10^{-26}~S_{e}^{~2.1}$ g C$_{60}$ per cm traversed in
C-rich rock.
Assuming a spherical rock geometry, with 
average trajectory length equal to its radius, the limit on their 
flux attainable by processing a rock mass $M$(kg) of 
age $t$(y) is then: 
\begin{eqnarray}
\Phi(cm^{-2}s^{-1}sr^{-1})<\mathrm{max}~[~8.2\times10^{3}
~\varepsilon/(\emph{M t S}_{e}^{~2.1})~,{}
\nonumber\\
{}8.3\times10^{-12}~\emph{M}^{-2/3}\emph{t}^{-1}~]
\end{eqnarray}
where $10^{-10}\varepsilon$~g is the minimum chromatographically-detectable 
C$_{60}$ mass ($\varepsilon_{\tiny{H\!P\!L\!C}}\!\!=\!\!1$).
The first term judges the ability of the particles to 
produce enough fullerenes during the lifetime of the 
sample, whereas the second imposes that at least 
one such particle must have crossed it (i.e., dictates the minimum 
exposure $Mt$ 
necessary for the first term to apply). For magnetic monopoles, this 
method is essentially applicable only to those with $\beta>10^{-1}$, 
for which the 
main mode of energy loss is via ionization. In this sense, this 
proposal is complementary to mica searches \cite{Beauford}, where the  
sensitivity decreases rapidly above $\beta\sim10^{-3}$. For $\beta$=1, 
monopole energy losses in C-rich rock depend on their magnetic charge 
$g\!=\!n~\!g_{D}$ via $S_{e}\simeq1.6\times10^{3}~n^{2}$\cite{Derkaoui} ($g_{D}$ 
= Dirac charge). Inspection of Eq.(4) 
shows that for 
$n > 3$,~$t\sim10^{9}$ and $M<50$, an improvement of the best current 
limits \cite{Orito} in this velocity range 
is already within reach. For $n=1,2$, a
decrease in $\varepsilon$ must be awaited: using a reasonable 
$M<$100, future or alternative
C$_{60}$ detection techniques can ameliorate the present experimental 
limits for GUT fast monopoles ($\Phi<3\times10^{-16}$) by several orders of 
magnitude before running into the limitations expressed by the second 
term in Eq.(4), with the attainable radiopurity of the rock as the 
only constraint. 

Coherent states of squarks and sleptons, predicted by supersymmetric 
generalizations of the standard model and dubbed $Q$-balls \cite{Coleman}, 
are also expected to be highly penetrating and to have a flux of 
$\sim\!\!1.25\times10^{3}\ Q^{-3/4}_{B} \left(\frac{1\ TeV}{m}
\right)$ cm$^{-2}$ s$^{-1}$ if 
they play a significant role as galactic dark matter
\cite{Kusenko}. Their mass, $m$, is assumed to be in the 0.1-100 TeV range, 
and their baryon number must respect $Q_{B}\!>\!10^{15} \left( m / 
1\ TeV \right)^{4}$. Their stopping power in matter of density $\rho$ should be 
$\sim\!\!10^{4} \frac{\rho}{1\ g/cm^{3}}$ keV/$\mu$m, but the energy loss 
mechanism depends on their being charged or neutral \cite{Kusenko}. Here 
it is conservatively assumed that only charged $Q$-balls are able to provide a 
dense enough energy transfer to form fullerenes. In that case, a modest 
reduction in $\varepsilon$ would bring about a challenge to the $Q$-ball sensitivity of large underground 
detectors like Super-Kamiokande (surface area =$7.5\times10^{7}\mathrm{cm}^{2}$).

Stable or metastable condensates of up, down, and strange quarks 
({\frenchspacing a.k.a. strange} quark matter or 
``nuclearites''), with mass numbers 
ranging from small nuclei to neutron stars ($A\sim10^{56}$), may constitute 
the ground state of hadronic matter \cite{Alvaro,Madsen}. If they 
make up an important 
fraction of the galactic dark matter halo, their local flux should be 
$\Phi_{S\!D\!M}\!=\!6\times10^{5}A^{-1}$\cite{Alvaro}. Their stopping power in rock
($4\times10^{-6}A^{2/3}(\frac{v}{300\mathrm{~km/s}})^{2}$ keV/$\mu$m
for $A>10^{15}$) is 
so enormous that the associated shock wave could in some cases create
a visible scar of melted material, an 
``astroblem''\cite{Alvaro} (for large 
$A$ the electronic 
and nuclear temperatures in the shock wave are assumed to balance out, making $S_{e}$ a 
sizable fraction of the total stopping power). An interesting 
peculiarity of the  method proposed here becomes evident in 
searches for nuclearites: Other techniques (mica, plastic track 
detectors, scintillators) rely on the principle ``one particle, one 
event'', and therefore their flux limits are blind to the value of 
$A$ (i.e., of $S_{e}$). 
In contrast to this, Eq.(4) improves as $A^{-2.1\times(2/3)}$ for 
nuclearites. Keeping in mind that $\Phi_{S\!D\!M}\!\propto\!A^{-1}$, 
this means that a negative C$_{60}$ search could in principle 
exclude nuclearites as the galactic dark matter for \emph{all} values 
$A\!>\!10^{13}$ ($A\!>\!10^{25}$ being not yet ruled out). Needless to say,
this ability is maimed by the second term in 
Eq.(4). All the same, such a search can immediately improve the 
existing sensitivity to these particles (a limit of $\Phi\!<\!10^{-20}$ for $A\!>\!10^{14}$) 
by two orders of 
magnitude after processing a realistic 
$M\!\sim$100. The possibility that lighter 
aggregates of strange matter (strangelets) may reach deeply into the 
atmosphere\cite{Wilk}, a hypothesis put forward 
to explain several anomalous (``Centauro'') events in detectors, can also be tested in this way
(strangelet stopping 
power in rock is expected to weight in at 
$\sim3.5\times10^{4}$ keV/$\mu$m \cite{Alvaro}).

In conclusion, an ubiquitous abundance of C$_{60}$ in rocks could be 
the signature of {\it any} new family of energetic highly-ionizing 
elementary particles. Its absence would allow in most cases to further 
push existing 
experimental limits, no small feat given their severity. 
In view of the incipience of studies of fullerene production by ion 
beams, our conservative estimates should be taken at face value (e.g., 
temperature- and material-dependent effects\cite{Bitensky,Brinkmalm} can increase the exponent
of $S_{e}$ in Eq.(4), resulting in an 
enhanced signal-to-(fission)noise ratio for this search). 
Irradiations of C-rich rocks to measure the production/recovery of fullerenes 
are then in order, 
together with searches for a natural fullerene 
presence correlated to spontaneous fission of $^{238}$U (possibly 
a new tool in geological dating).\\
{\it Acknowledgements:} 
We are indebted to P.B. Price for a critical reading of the manuscript.

\end{document}